\documentclass[preprint,3p,twocolumn]{elsarticle}

\usepackage{lineno,hyperref}
\modulolinenumbers[5]

\journal{Nuclear Physics B}

\usepackage{graphicx}
\usepackage{dcolumn}
\usepackage{bm}
\usepackage{amssymb}
\usepackage{amsthm}
\usepackage{amsmath}
\usepackage{caption}
\usepackage{setspace}
\usepackage{enumitem}
\usepackage{textcomp}
\usepackage[section]{placeins}
\usepackage[utf8]{inputenc}
\usepackage{wrapfig}
\usepackage{epstopdf}
\usepackage[toc,page]{appendix}
\usepackage{enumitem}
\usepackage{xcolor}

\newcommand{\beq}{\begin{equation}}
\newcommand{\eeq}{\end{equation}}

\begin{document}

\begin{frontmatter}

\title{Asymptotic safety and quantum gravity amplitudes}
\author{Jan H. Kwapisz\corref{mycorrespondingauthor}}
\ead{Jan.Kwapisz@fuw.edu.pl}
\author{Krzysztof A. Meissner}
\address{Faculty of Physics, University of Warsaw,
ul. Pasteura 5, 02-093 Warsaw, Poland}
\ead{Krzysztof.Meissner@fuw.edu.pl}

\date{\today}

\begin{abstract}
Application of the Weinberg's conditions of asymptotic safety to amplitudes and not to couplings of the effective action can help to uniquely define the running in the UV and avoid some of the asymptotic safety program problems. As it turns out the requirement for a quantum gravity to be defined by S-matrix, that obeys the Weinbergs asymptotic safety criteria is a very restrictive one. The idea is illustrated by the 4-graviton amplitude in string theory and its symmetries.
\end{abstract}
   
\begin{keyword}                         
Asymptotic safety \sep renormalisation group \sep string theory \sep graviton interaction\\
\emph{PACS}: 04.60.-m 11.25.-w 14.80.Bn
\end{keyword}                           

\end{frontmatter}

\section{Introduction.}

In physics there are four known fundamental interactions: strong, weak, electromagnetic and gravitational. The first three can be quantised by means of perturbative quantum field theory approach and these theories are predictive since a finite number of experiments is sufficient to fix all of the couplings in the action. On the other hand if one quantises gravity in a similar way, one gets a theory which requires infinite number of experiments to specify its predictions. Therefore one has to find a new way to quantise gravity to make it predictive again.

In order to do so one can either relax (some of the) assumptions of general relativity and/or quantum theory in order to make it predictive or propose a new quantisation scheme (non-perturbative one). In the first category there are for example higher curvature gravity theories \cite{PhysRevD.16.953,Voronov:1984kq,Anselmi:2018tmf,Donoghue:2018lmc,Donoghue:2019fcb} or Ho{\v{r}}ava-Lifschitz gravity \cite{Horava:2009uw}. There are also discrete spacetime approaches such as loop quantum gravity \cite{PhysRevD.36.1587,Rovelli:1989za} or causal dynamical triangulations \cite{Ambjorn:2001cv,Ambjorn:2002gr}. In string theory one abandons the whole concept of quantum field theory, see for example \cite{green_schwarz_witten_2012,polchinski_1998} and \cite{Green:1984sg,Gross:1984dd,Candelas:1985en}.

On the other hand one can accept that gravity indeed can possess an infinite number of couplings, but the theory is nevertheless predictive as long as the values of (almost) all couplings can be calculated theoretically and they are functions of only finite number of experimental values. This is for example the case of asymptotic safety in quantum gravity (ASQG) \cite{Weinberg,WeinbergAS,QFTWeinberg}, where one requires that all of the couplings reach a (non)-trivial fixed point in the UV, which fixes almost all the couplings. Using the Functional Renormalisation Group (FRG) \cite{Wetterich:1992yh,Morris:1993qb} techniques such a tentative fixed point was found \cite{Reuter:1996cp,Souma:1999at,Lauscher:2001ya,Reuter:2001ag} (previously conjectured by other considerations  \cite{Gastmans:1977ad,Christensen:1978sc,Jack:1990ey,Kawai:1995ju,Smolin:1981rm}). Furthermore this approach gained much attention in the recent years due to more compelling evidence for the fixed point \cite{Gies:2016con,Eichhorn:2018yfc} and rich phenomenological behaviour, such as prediction of Higgs mass in Standard Model \cite{ShaposWetterich} and in various extensions \cite{Grabowski:2018fjj,Kwapisz:2019wrl}, prediction of masses of various quarks \cite{Eichorntop,Eichornquarks} and properties of dark matter \cite{Reichert:2019car,Hamada:2020vnf}.

Despite these successes, recently the whole programme was challenged by some issues \cite{Donoghue:2019clr} of fundamental importance, see also \cite{Anber:2011ut} and \cite{Bonanno:2020bil}. Namely if one considers the quantum gravity in the effective field theory (EFT) approach below Planck scale, then due to lack of \emph{crossing} and \emph{universality} properties in the amplitudes \cite{Donoghue:2019clr,Anber:2011ut} one cannot introduce a single running of the Newtonian coupling $G_N(\mu)$ to describe their energy dependence. It also happens in the pion model, which is the EFT for quantum chromodynamics, where the pion amplitudes doesn't posses the crossing symmetry \cite{Chivukula:1992gi,Cavalcante:2001yw} in the Mandelstam variables (not the crossing LSZ symmetry: $s->u$, which is the one discussed in the article and required for unitarity). On the other hand the QCD amplitudes does have in the massless (high energy) limit. In this letter we discuss this problem and propose a possible cure. The idea is to use a specific amplitude related to the coupling we are interested in, try to find a formula valid at all scales, and only then ask a question about the asymptotic safety. Such an approach requires a reformulation of the Weinberg's criteria of asymptotic safety. In what follows we concentrate on the 4-graviton amplitude which is directly related to the Newtonian coupling. To have a well defined amplitude for all energies without encountering the usual difficulties we have chosen to use the known form of this amplitude calculated within the framework of string theory, which is finite also at the loop level, to show that the concrete definition proposed in this letter can have an explicit realization and can solve also other issues posed in \cite{Donoghue:2019clr} (we would like to emphasize that we don't claim that the string theory is {\it the} theory of quantum gravity, but we use it as an illustration since it is the only framework existing at present to consistently calculate the graviton amplitudes). 

\section{Running of the couplings.} In order to understand the criticism expressed in \cite{Donoghue:2019clr} we review the notion of running couplings in the quantum field theories (QFT). 
In experimental physics one investigates the properties of matter by colliding the particles prepared in a given state (momenta etc), measure the outcome of the collisions. and compare it with the theoretical amplitudes calculated in the QFT framework. 
These amplitudes are encoded in the $S$-matrix, which is an array of probabilities between all the possible non-interacting in-states in the far past and all the possible non-interacting out-states in the far future. Tree level values of amplitudes get (infinitely) corrected by loop diagrams. Therefore, to have correct definitions of in/out states, such that they are not-interacting, one redefines (renormalizes) the splitting of the Hamiltonian into free and interaction parts, see \cite{QFTWeinberg,Peskin:1995ev}. Then all the infinities are absent in the amplitudes.  The renormalisation procedure introduces another source of scale (energy) dependence into the theory on top of the classical one by introducing counter-terms to absorb the infinities and defining the theory in terms of renormalized couplings $g_R=g_B-\delta g$, where $g_B$ are the bare couplings.

In principle one should write all possible counter-terms respecting the symmetries of the theory, however if a classical Lagrangian has only couplings of non-negative dimension $\Delta_i \geq 0$, then by power counting arguments one can show that only the $n$-point interactions with $n\leq 4$ can be divergent. In that case the counterterms will have the same structure as the original terms of the bare lagrangian. 

This fact has two further consequences. First, for dimensionless couplings (we neglect the masses in our discussion due to near conformal structure of the Standard Model \cite{Meissner:2006zh,Meissner:2007xv,Chankowski:2014fva,Latosinski:2015pba,Lewandowski:2017wov,Kwapisz:2017vjt,Grabowski:2018fjj}) the corrections are proportional to the couplings itself and the sign of the correction will not be channel dependent (\emph{crossing}). Furthermore the 4-point diagrams (interactions with 4 external legs) exhibit the crossing symmetry, i.e. permutation symmetry in the Mandelstam variables $s,t,u$ \cite{Mandelstam:1958xc} (\emph{universality}). These two facts mean that the divergent and finite parts of the amplitudes can be described by the same, universal behaviour \cite{Donoghue:2019clr,Anber:2011ut,PhysRevD.2.1541,Symanzik:1970rt,Symanzik:1971vw}. 
Then the entire momenta dependence of the theory is contained in a couple of energy dependent functions which in case of perturbatively renormalisable theories are exactly couplings in the Lagrangian. Then the scale dependence of the coupling is given by the renormalisation group equations:
 \beq
 \label{betafunction}
 \mu \frac{\partial g_i}{\partial\mu} = \beta_i.
 \eeq
 Obviously this change of couplings with scale is not physical in itself and the coupling has to be tied to observable quantities at each scale. The smaller the coupling the better the process is described by the tree-level diagrams only. Standard Model possesses only marginal couplings and one relevant (Higgs mass) and can be quantised using the quantum field theory loop expansion due to the couplings lying in the perturbative range.
 
\section{Quantising General Relativity.} On the other hand the action of General Relativity is given by
\beq
\label{EHaction}
S= \frac{1}{8\pi G_N}\int d^4 x \sqrt{-g} R,
\eeq
where the dimension of $[R]=2$, then $[G_N]=-2$. By the power-counting argument the quantum gravity is the theory with infinite number of counterterms (if there are no cancellations as in supergravity \cite{deWit:1982bul,Kallosh:1980fi,Kallosh:2010kk}), however one cannot make infinite number of experiments to determine the predictions of the theory. The lack of cancellations was confirmed by the explicit Feynman diagrams calculations showing that General Relativity requires counterterms not proportional to $R$ at one-loop level when coupled to matter\cite{tHooft:1974toh,Christensen:1979iy} (which disappear in the matter free case by the equations of motions) and at two loops in the matter free case \cite{Goroff:1985th,vandeVen:1991gw}. 

\section{Asymptotic safety programme.} Gravity treated as quantum field theory has infinite number of (counter)-terms in Lagrangian. However this doesn't necessary mean that the theory is un-predictive at arbitrary scales. After all, the quantum effective action of $\phi^4$ theory also possesses an infinite number of terms but the value of all the couplings can be calculated from two quantities $\lambda_{ph}$ and $m_{ph}$, closely related to observations. Steven Weinberg \cite{Weinberg,WeinbergAS} noticed that quantum gravity (QG) could be predictive if all the irrelevant ($\Theta_i<0$) couplings could be determined in terms of the relevant and marginal ones. The so called asymptotic safety mechanism could be operative if all the couplings of the theory reached a fixed point (such that the right hand side of Eq.~\ref{betafunction} vanishes) and then the theory will not diverge when moving with the scale towards UV \cite{Weinberg,WeinbergAS,Shore:2016xor} and get closer to conformal invariance. If all the couplings of a theory reach zero at the fixed point then we call the theory asymptotically free, otherwise when $g_i^{\ast} \neq 0$ we call it asymptotically safe. In particular if the gravity is asymptotically safe, then the dimensionless $\tilde{G}_N = G_N \mu^2$ should have a fixed point. 

In the early days the running of $G_N$ was found in $2$-dimensions and its vicinity \cite{Gastmans:1977ad,Christensen:1978sc,Jack:1990ey,Kawai:1995ju} by means of the $2+\epsilon$ dimensional expansion. Similar claims were also made with the $1/N$ expansion approach \cite{Smolin:1981rm}. However only after applying the Wilsonian inspired exact Wetterich-Morris-Ellwanger equation  \cite{Wetterich:1992yh,Morris:1993qb,Ellwanger:1993mw} to tackle this problem the approximate fixed point was found in seminal Reuter article \cite{Reuter:1996cp,Lauscher:2001ya,Reuter:2001ag} and then explored in various truncations. 

Applying this equation to quantum gravity, even though it is exact, raises many problems of both fundamental and technical type, see \cite{Donoghue:2019clr,Eichhorn:2018yfc,Branchina:2003ek,Wetterich:2019qzx,Bonanno:2020bil}, but one could hope that even problematic, these issues can be solved in the future. However, recently in \cite{Donoghue:2019clr,Anber:2011ut} the whole notion of the running gravitational coupling was challenged. The argument is the following. The running coupling is merely a description of a collective behaviour of amplitudes and should be reflected in them. Since $[G_N]=-2$ the corrections to the amplitudes should be proportional to $G q^2$, where $q^2$ is one of the Mandelstam variables \cite{Dunbar:1994bn,Anber:2011ut}, see also \cite{PhysRevLett.66.1669,Bern:1993wt}. Since in different reactions $q^2$ can be positive or negative, then depending on the reaction the gravitational correction can go into different directions. Therefore the notion of the universal running cannot be introduced for the Newtonian coupling in four dimensions.

\section{Running in terms of amplitudes.} The asymptotic safety criteria are very appealing theoretically for testing the consistency of the theory. For this purpose we can rephrase them as follows: the theory should approach the scale invariant regime in the UV, requires finite number of experimental input to be fully predictive and all the observables should be finite. Furthermore Weinberg conjectures that approaching the UV limit should be universal and channel independent, hence (in terms of gravity) we only consider $G_N(\mu)$ rather than $G_N(p_i)$. 

The amplitudes are the main observables in QFT (by their direct connection to the cross sections) therefore we propose to identify the ``running'' with the change of the amplitude with scale and not of any particular term in the (effective) lagrangian.

The importance of the amplitudes approach rather than the effective action treatment can be clearly seen in string theory. In string theory one can calculate the low-energy effective action using the the non-linear sigma model approach \cite{PhysRevLett.45.1057,Callan:1985ia}, but the string theory amplitudes cannot be reproduced by any of this effective actions since finiteness of string amplitudes is tied to the modular invariance that cannot be reproduced by means of a lagrangian of any conventional QFT. 

One should note that there may be other symmetries of the amplitude (and of the corresponding effective action) like $O(d,d)$ symmetry in string theory \cite{Meissner:1991ge,Meissner:1991zj} that tie together many terms in the lagrangian -- for example a 4-graviton amplitude involves many higher curvature terms that are uniquely fixed by the symmetry \cite{Meissner:1996sa,Kaloper:1997ux}.

Motivated by the original Weinberg notion of asymptotic safety we propose three criteria in terms of UV behaviour of amplitudes. First, specifying finite number of amplitudes is sufficient to derive all of the theory predictions. Second, all the amplitudes should possess scale invariant regime for very large energies (near the cutoff) \cite{Gross:1970tb}:
 \beq
A(p_1,p_2,\ldots) = \lambda^{-d} A(\lambda p_1, \lambda p_2, \ldots),
 \eeq
 where $d$ is the dimension of $A$. In terms of quantum field theory this corresponds to the massless limit. Third, we demand the $E\to +\infty$ limit to be unambiguous for the 'running' part of amplitude $A'$, i.e. the amplitude without the polarizations' contractions. In particular for the 4 identical particle elastic scattering this means that 
 \beq
\lim_{s\to \infty, t-\textrm{fixed}} A'(s,t,u)= \lim_{t\to \infty,  s-\textrm{fixed}} A'(s,t,u).
 \eeq
We call this asymptotic crossing symmetry. In particular first and third conditions ensure finiteness and UV-fundamentality of the theory and the second condition ensures the scale invariance of the UV-phase. Furthermore if we do not assume the asymptotic crossing symmetry then there is no universal notion of running governing the high energy behaviour of the theory. It is no longer described by any universal “running” and there is no notion of fixed points of this running, which was the main objection of \cite{Donoghue:2019clr}. In the context of quantum gravity one should require that the theory has a graviton in its spectrum and the four graviton amplitude is not trivial. Another reason for the asymptotic crossing symmetry is that then the low energy limit is described by a single coupling $G_N$, which we then match to high energy physics. The amplitude defined in that way satisfies both \emph{crossing} and \emph{universality} properties by construction.
Given these criteria we identify the ``running'' in quantum gravity with the change of 'scalar' part (multiplying the polarization tensors kinematics part) of the 4-graviton amplitude in the full theory. Finally let us note that in this picture there is no notion of problematic Wick rotation because we do not have to calculate the amplitude in the Euclidian signature and then analytically continue into Lorentz one. Finally unitarity is manifest by the optical theorem and correct structure of the poles. As we discuss below string theory in the gravitational sector satisfies our criteria. 

\section{String theory amplitudes example.} Actually the crossing symmetry was the main motivation for the Veneziano amplitude \cite{Veneziano:1968yb}, which gave birth to the string theory \cite{1970sqmconf269N}. In Type II superstring theory the Virasoro-Shapiro amplitude (closed string analog for Veneziano amplitude for four gravitons) takes the following form \cite{green_schwarz_witten_2012}
\beq
A(p_1,p_2,p_3,p_4)= 2 g_D^2K_{cl} C(s,t,u),
\eeq
where $C(s,t,u)$ is given by
\beq
C(s,t,u)= -\pi\frac{\Gamma(-s/8)\Gamma(-t/8)\Gamma(-u/8)}{\Gamma\left(1+\frac{s}{8}\right)\Gamma\left(1+\frac{t}{8}\right)\Gamma\left(1+\frac{u}{8}\right)}
\eeq
and $K_{cl}$ is a lengthy kinematic (cross-symmetric) and polarization factor polynomial in $s,t,u$. If we vary $s$ and keep $t$ fixed, then this amplitudes has infinite number of simple poles \cite{Tong:2009np} on a real line at $s= 8n$, with $n\geq0$. This can be understood as summing over infinite number of particles with growing masses in the $s$-channel. The similar conclusion can be drawn if we keep $s$ fixed and vary $t$. This property is called duality. Hence we cannot take large energy limit ($s\to \infty$, $s/t$ fixed) along the real line so we shift $s$ slightly in the imaginary direction. Ultimately, the poles are anyway shifted off the real axis making the $s\to \infty$ limit well defined because massive states are unstable in the interacting theory, hence their amplitudes possess non-trivial imaginary part. As a result we get the exponential fall-off:
\beq
C(s,t,u) \varpropto \exp\left(-\frac{1}{4} (s\log s+ t\log t + u\log u)\right),
\label{amplgrav}
\eeq
satisfying our criteria. Our conditions are also satisfied beyond the tree level amplitude, since this behavior is also present for the resumed loop amplitude in this limit \cite{Amati:1987wq,Gross:1987ar,Amati:1987uf} (string theory preserves generally the crossing symmetry at the loop level \cite{deLacroix:2018tml}). On the other hand in the low-energy limit the Veneziano amplitude can be matched with tree-level graviton amplitude, since $C(s,t,u) \sim \frac{1}{stu}$ with $\kappa = \frac{1}{2} \alpha' g^2$ for heterotic strings \cite{green_schwarz_witten_2012}, where $g \approx 1$ and $\alpha' \approx M_P^2$. Furthermore the second term in the expansion of the Virasoro-Shapiro amplitude has the opposite sign as the first term (in the expansion of the beta function) which is a hint for asymptotically safe anti-screening behaviour as an intermediate step between classical gravity and string theory \cite{deAlwis:2019aud} but the issue deserves further investigation. In the future work we are planning to check whether the bounds discussed in \cite{Adams:2006sv} are satisfied and hence whether GR EFT can be UV completed by string theory.

It should be emphasized that the fact that the amplitude (\ref{amplgrav}) asymptotically vanishes does not say anything about the individual contributions from the interaction vertices in the lagrangian, in particular whether the Newton's constant $G_N$ asymptotically vanishes or not. However, this asymptotic behavior points to an important and far reaching change of perspective with respect to the usual approach where only the Newton's constant and its running are taken into account.

\section{Remarks} Before coming to the conclusions let us make a few remarks.
Our approach can be complementary to the other studies of asymptotic safety phenomena such as vertex expansion \cite{Christiansen:2012rx,Christiansen:2015rva,Meibohm:2015twa,Denz:2016qks} of the $\Gamma_k$ or causal dynamical triangulations \cite{Ambjorn:2001cv,Ambjorn:2002gr}, yet the crossing symmetry is not manifest there. It would be interesting to compare those in the future. On the other hand the recent proposals of redefining the FRG approach to quantum gravity \cite{Knorr:2018kog,Knorr:2019atm} puts the whole momentum dependence into the the form factors for higher curvature terms in the lagrangian. Actually the amplitudes for scalar particles scattering including the graviton propagators were recently calculated and they also satisfy our criteria \cite{Draper:2020bop,Draper:2020knh}. However the coefficients of $\Lambda$ and $R$ doesn't run in any sense in that approach, since the form factors are attached to the $C^2$ and $R^2$ terms in the effective action. This can make the physical running of higher curvature couplings essential to the Weinberg proposal, rather than the Einstein-Hilbert truncation see \cite{PhysRevD.16.953,Voronov:1984kq,Anselmi:2018tmf,Donoghue:2018lmc,Donoghue:2019fcb}. Furthermore due to lack of crossing symmetry in the crucial 4-graviton amplitude in this approach \cite{Donoghue:2019fcb,Draper:2020knh} one cannot also rely on the RG-improvement and identify the "running" scale of the Wetterich equation with some physical scale, which is usually done. These recent developments suggests that maybe some multi-scale generalisation of Wetterich equation could be in order to capture the dependence of the amplitudes, but it seems a long way ahead.

In this letter we focus only on the amplitudes with gravitons on external lines, since running of the Newton's constant is the crucial issue in the whole AS scenario. If we include matter there will be obviously other amplitudes with gravitons. Yet with a proper treatment of the polarisation contractions (on top of the kinematical part) the whole amplitude can still satisfy the \emph{crossing} and \emph{universality} properties. The argument behind this conjecture relies on the double copy 'equivalence' (for $N>4$ supergravities) gravity and Yang-Mills squared \cite{Bern:2010ue} if we properly identify gauge with polarisation contractions and gauge with gravitational couplings. In the string theory the situation is much clearer due to the KLT relations \cite{Kawai:1985xq} and the linear relations between amplitudes conjectured by Gross and Mende \cite{Gross:1987kza,Gross:1988sj,Gross:1988ue} (for proof see \cite{Lee:2015wwa} and references therein) where all the closed strings amplitudes stripped from kinematics share the Gamma function behaviour and posses an exponential fall-off \cite{Chan:2004tb,Chan:2006qj} at tree level. This actually allow for a single coupling running interpretation in context of all the interaction in the string theory (at least at tree level).

On the other hand it is conjectured that the high energy gravitons scatterings are dominated by the productions of black holes, see \cite{Giddings:2010pp,Giddings:2011xs}. Yet it raises a question whether such an amplitude can be tested and not whether it exists at all, which is a slightly different question than the one we are answering. Actually this a question for a non-perturbative definition of quantum gravity, a programme which didn't succeed ultimately in any approach. The closest to such a formulation is an AdS/CFT conjecture \cite{Maldacena:1997re} which is however incomplete from the theoretical point of view \cite{Banks:2003vp,Banks:2019oiz} and works only in AdS spacetime. Yet due to the anti-screening character of the asymptotic safety one can hope that these issues can be resolved in our language and we leave that for future work, see also a recent resolution of that problem proposed in \cite{Addazi:2020nkm}, and the study of this problem in Ho{$\rm\check{r}$}ava gravity \cite{Elizalde:2011tx}.

\section{Conclusions.} Four lessons can be learnt from our investigation. First, we emphasized the importance of crossing symmetry in the Weinberg's asymptotic safety proposal (with a concrete example of string theory graviton amplitudes). Second, we propose to associate the running with the amplitudes and not with couplings in the lagrangian (which is actually done in the conventional QFTs since there both notions coincide) - they have very definite UV-behaviour avoiding the ambiguities in the running emphasized in \cite{Donoghue:2019clr}. It seems that if one wants to apply Weinbergs criteria and requires physical, single-scale running of the 4-graviton amplitude such that in low energy it is tied to $G_N$ is a very restrictive criteria. This fact points to an important conclusion that the running of the Newton's constant is not sufficient to predict or model the UV behavior of gravity and the other degrees of freedom may play an important or even decisive role.

Thirdly our criteria shows the importance of matter-gravitational interactions and their study in terms of quantum gravity. For example the amplitudes for pure gravity lack the crossing symmetry at one-loop \cite{Dunbar:1994bn} and two-loop level \cite{Abreu:2020lyk}. On the other hand the $N=8$ supergravity amplitudes have the crossing symmetry \cite{Dunbar:1994bn} also at higher loops \cite{Henn:2019rgj}, which is by the way another example satisfying our criteria. Furthermore the result  \cite{Dunbar:1994bn} still has to be unitarised and it was shown \cite{Camanho:2014apa} that under certain assumptions in order to preserve causality one requires an infinite tower of higher spins states, such that they form the Veneziano amplitude and the EFT result in UV completed by string theory. This might be the hint that Einstein-Hilbert truncation is only the low energy approximation and hence $L\sim \frac{1}{G_N}(R+\ldots)$ and only after taking the dotted part into account one can introduce the meaningful ``running ''of $G_N$ as the non-linear sigma models hints. It would be extremely interesting if such Lagrangian could be obtained without the supersymmetric degrees of freedom and without a tachyon. Let us also note that the amplitudes of linear-sigma models satisfy our claims in the non-gravitational context \cite{Adams:2006sv} and have a non-vanishing value in the UV.
Also using the amplitudes picture one can check the effects of quantum gravity interactions on matter and understand the asymptotic safety conditions on the particle properties in the Standard Model and beyond \cite{ShaposWetterich,Grabowski:2018fjj,Kwapisz:2019wrl,Eichorntop,Eichornquarks,Reichert:2019car,Hamada:2020vnf,Dona:2013qba,Meibohm:2015twa} (one could also check whether this conditions agree or disagree with the stringy inspired swampland conditions \cite{Vafa:2005ui,Ooguri:2006in,ArkaniHamed:2006dz,Obied:2018sgi,Agrawal:2018own}). 

Finally the $N \geq 5$ supergravities and string theory have a vanishing conformal and chiral anomalies \cite{Meissner:2016onk,Meissner:2017qwm,Kallosh:2016xnm}, which allows for a conformal UV limit. On the other hand the effects of this anomalies can be disastrous on the Early Universe cosmology \cite{Godazgar:2016swl}. Hence it seems that there is a connection between asymptotic crossing symmetry (and asymptotic safety), finiteness of the amplitudes in 4 dimensions \cite{Freedman:2018mrv,Herrmann:2018dja,Abreu:2019rpt} and absence of conformal anomaly. It would be extremely interesting to check whether amplitudes calculated in the framework of the proposed $E_{10}$ symmetry \cite{Damour:2002cu,Damour:2006xu,Kleinschmidt:2014uwa,Meissner:2018gtx} also satisfy the criteria proposed in this paper.

\section*{Acknowledgements} We thank P. Chankowski, J. Donoghue, S. Giddings, S. Melville, H. Nicolai and M. Reichert for valuable discussions and for comments on the manuscript. K.A.M. and J.H.K. were partially supported by the Polish National Science Center grant DEC-2017/25/B/ST2/00165. J.H.K. was supported by the Polish National Science Centre grant 2018/29/N/ST2/01743.

\addcontentsline{toc}{section}{The Bibliography}
\bibliography{mybibfile}{}
\bibliographystyle{elsarticle-num}
\end{document}